
\input topp.tex

\PHYSREV
\tolerance 2000
\nopubblock
\titlepage
\title{ENERGY DECAY IN BURGERS TURBULENCE AND INTERFACE GROWTH. THE PROBLEM OF
RANDOM INITIAL CONDITIONS - II.}

\author{Sergei E. Esipov}

\address{James Franck Institute and Department of Physics, University of
Chicago, Chicago, Illinois 60637, USA}
\andaddress{Department of Physics and Material Research Laboratory,
University of Illinois at Urbana-Champaign,
1110 West Green Street, Urbana, Il. 61801-3090, USA}
\vfil
\endpage

\abstract{We present
a study of the Burgers equation in one and two dimensions $d=1,2$
following
the analytic approach indicated in the previous paper I.
For the problem of initial
condition decay we consider two classes of initial condition
distributions $Q_{1,2} \sim \exp\big[-(1/4D)\int({\nabla}h)^2$d{\bf x}\big]
where $h$-field is unbounded ($Q_1$) or bounded ($Q_2, |h|\leq H$).
In one dimension these distributions give examples of
non-degenerate and degenerate Burgers models of turbulence,
respectively.
Avoiding the replica trick and
using an integral representation of the logarithm
we study the exact analytically
tractable field theory which has $d=2$ as a critical dimension.
It is shown that the degenerate one-dimensional case has three
stages of decay, when the kinetic energy density diminishes with time as
$t^{-2/3}$, $t^{-2}$, $t^{-3/2}$ contrary to the predictions of the
similarity hypothesis based on the second-order correlator of the distribution.
In two dimensions we
find the kinetic energy
decay which is proportional to $t^{-1}\ln^{-1/2}(t)$.
It is shown that the pure
diffusion equation with the $Q_2$-type initial condition has non-trivial
energy decay exponents indicating connection with
the $O(2)$ non-linear $\sigma$-model.
}
\bigskip\noindent
PACS Numbers: 05.40.+j, 47.10.+g, 68.10.Jy

\endpage

The history of Burgers equation $^2$ shows that
the connection between Burgers turbulence and Navier-Stokes turbulence
is a complicated issue. Citing the recent paper by Gotoh and Kraichnan $^3$,
``by now it is clear that the differences between Burgers dynamics and
NS dynamics are at least as significant as the similarities.'' This
statement is made with respect to the one-dimensional version of
Burgers equation.
It is
also clear that Burgers equation above one dimension has
smaller relevance to real turbulence,
for instance, the kinetic energy
is not conserved in the invisid limit. In its turn, Burgers equation,
being the simplest
diffusive non-linear equation,
have a number of other physical applications to date. We may refer to
the comparative usage in RG study of NS equations $^4$,
the field of interface growth in the framework
of KPZ equation $^5$,
models of the large-scale structure of the Universe
$^6$, solid state physics applications $^7$, etc. As
in the NS turbulence there are different means to excite the
turbulent behavior. We think that a simpler problem to study is the
relaxation of random initial condition rather than random stirring
(external noise) problem.

The present paper is a continuation of our recent work on Burgers turbulence
and interface growth $^1$. In Ref.1 we have solved the problem for discrete
initial conditions and indicated a field-theoretical approach suitable for
continuous distributions. The field theory for Burgers equation with
continuous initial conditions resembles
the so-called Liouville model $^8$ of strings and was applied to
the non-bounded
gaussian distribution of initial conditions in one
dimension. Studying the corresponding Schrodinger equation we essentially
reproduced the Burgers result for the kinetic energy decay $^2$, $E(t) \sim
t^{-2/3}$.

The present work is partly
devoted to the study of {\it degenerate} Burgers turbulence in
one dimension,
and also contains an investigation of
the two dimensional case. We only study the decay of kinetic energy, the
simplest local correlator.
Our results
in degenerate $d=1$ case are in disagreement with the previously reported
$E(t) \sim t^{-1}$
transients $^{6,9}$ which derivation was based on the so-called similarity
hypothesis involving the second-order correlator (Loitsyanskii correlator)
of the distribution.
The explanation
is that the similarity hypothesis is based on dimensional arguments,
and in the degenerate case the Loitsyanskii correlator is zero by definition;
whereas it is impossible to describe initial conditions without any parameter.
With the distribution $Q_2$ we find that energy decays as $E(t) \sim t^{-2/3}$
as in the non-degenrate regime until the degeneracy becomes
important at some crossover time.
After that time the decay proceeds faster,
$E(t) \sim t^{-2}$ for some time and (if viscosity is not equal to zero
exactly) has a second crossover to the
purely diffusive
decay, $E(t) \sim t^{-3/2}$.

Studying the Burgers equation it is natural to compare its behavior with and
distinguish it from the pure diffusion. This implies that for
any of the cases considered it is the diffusion equation to be solved
first. Consequently, as a by-product of the method of Green functionals, we
present the study of the diffusion equation in $d=2$ for the bounded
distribution $Q_2$. The exponent for the energy-decay power law
has a continuous dependence on the parameter $D/H^2$ (see below) up to
a critical value $D/H^2=8/\pi$ when the exponent stops to vary. This
observation implies some connection with
the $O(2)$ non-linear $\sigma$-model $^{10,11}$.

The Burgers equation in two dimensions leads to functional equations for
Green functionals. The invisid limit can be solved and the results show
explicit dependence upon short-distance cut-off. The kinetic energy decays
as $E(t) \sim t^{-1}\ln^{-1/2}(t)$ for the $Q_1$-distribution. The onset
of ``boundness'' is manifested by a sudden drop of the kinetic energy to
zero (in the invisid limit), thus showing some resemblance to the
one-dimensional case. The reported results are beyond the reach of scaling
arguments $^{12}$ which are helpful in the case of
single power-law
dependences.

\chapter {Burgers equation as a field theory}

We briefly re-derive the field theory for the
problem of Burgers equation with random initial conditions in $d$-dimensions
for the sake of completness.
Consider the Burgers equation $^2$
$$\partial_t {\bf v} = \nu \nabla^2 {\bf v} - {1\over2}
{\bf \nabla}v^2,\eqn\burg$$
which with the help of the
velocity potential ${\bf \nabla}h=-{\lambda^{-1}}{\bf v}$
gives an equation
$$\partial_t h = \nu \nabla^2 h + {\lambda\over2}({\bf \nabla}h)^2,\eqn\kpz$$
also known as the deterministic KPZ equation $^5$.
The parameter $\lambda$ is introduced for convenience, it has
dimensionality of lenght over time,
and hencefore $h$ has the dimensionality of lenght.
The Hopf-Cole transformation to the new function $\exp({\lambda}h/2\nu)$
results in a diffusion equation which is being solved for a given
(transformed) initial condition. The solution in terms of some initial $h_o$
reads
$$h({\bf x},t) = {2\nu\over\lambda}\ln \int d{\bf y}g({\bf x}-{\bf y},t)
\exp {\lambda\over2\nu}h_{o}({\bf y}) \eqn\sol $$
where $g({\bf x},t)=(4\pi\nu t)^{-{d\over2}}\exp[-(x^{2}/4\nu t)]$ is the
heat kernel (Green function of the diffusion equation).
The kinetic energy density that we study here is defined as
$E(t)={\lambda^2\over2}\langle({\bf \nabla}h)^2\rangle =
\lambda\partial _{t}\langle h \rangle$, where we used translational
invariance to drop the diffusion term. Using an integral representation
of the logarithm in \sol\ we obtain
$$\langle h \rangle = {2\nu\over{\lambda}} \int\limits _{0}^{\infty} {du\over
u}
[e^{-u} - \psi (u,t)],\eqn\heig$$
and
$$E(t)=2\nu\partial _{t}\int\limits _{0}^{\infty} {du\over u}
[e^{-u} - \psi (u,t)] \eqn\energy $$
where
$$\psi (u,t) = \int {\cal D}[h_{o}] \exp \{-S[h_{o};u,t]\} \eqn\pathint .$$
This is written in the form of a field theory defined by the following action:
$$S[h_{o};u,t] = \int d^{d}x \Bigl \lbrace {1\over 4D}(\nabla h_{o})^{2}
+ug({\bf x},t)e^{\alpha h_{o}({\bf x})} \Bigr \rbrace .\eqn\liou $$
It resembles closely the Liouville model in string theory but differs
by the heat kernel $g$, which makes the potential to be $x$-dependent
(i.e. ``time''-dependent). The one-dimensional case of this theory was
considered in Ref.1.

\chapter{Degeneracy of Burgers turbulence}

We are interested in different
classes of distribution of random initial conditions which evolve unlike
each other. We have had such an example in the case of discrete initial
condition studied in Ref.1. In this Section we present a comparison between
bounded and unbounded gaussian distributions of initial conditions and
relate them to the degeneracy of the Burgers turbulence in one dimension.

The degeneracy of turbulence is related to the Loitsyanskii correlator, $D'$.
By analogy with the Navier-Stokes turbulence the Loitsyanskii correlator
(in one dimension) is defined as $^{13,2,6,9}$
$$D' = {1\over{2\lambda^2}} \int_{-\infty}^{\infty}dx
\langle v(x)v(0)\rangle, \eqn\loi$$
It is known that the
physics of (Burgers) turbulence is different depending on whether this
correlator is zero or not. Usage of the velocity potential $h$ makes it
easier to understand the physical sense of $D'$. Indeed, take into
account the
adopted gaussian initial conditions $Q_{1,2}$ given by
$$Q[h_o] \sim \exp\Big[-{1\over{4D}}
\int (\nabla h_{o})^{2} d^{d}x\Big] \eqn\grads $$
and consider the correlation of the velocity
potential $h$ following Burgers $^2$.
We have
$h_o(x) - h_o(0) = - \lambda^{-1}\int_0^x v dx.$
The two-point correlator of interest is given by
$$\langle (h_o(x) - h_o(0))^2 \rangle = \lambda^{-2}\int_0^x\int_0^x v(x')
v(x'') dx' dx'' =
$$
$$= 2\lambda^{-2}
\int_0^x dz (x-z)\langle  v(z)v(0)\rangle = 2D'x + O(1).\eqn\hsq$$
Consequently all even correlators are
$\langle (h_o(x) - h_o(0))^{2n}\rangle =
(2n-1)!!(2D'x)^n + O(x^{n-1})$ and Burgers concludes that the distribution
of $h$ at a given distance $x$ is gaussian, $P(h_2,h_1,x)=(4\pi{D'}|x|)^{-1/2}
\exp[{-(h_2 - h_1)^2/4D'|x|}]$. This speculation relies only on the
fact that the problem is translationally invariant and
Loitsyanskii correlator $D'$ is non-zero, while its support is finite.
We can see now
that the non-bounded and bounded $h$-distributions \grads\ belong to
non-zero and zero $D'$-distributions, respectively, and thus
give representative
examples of degenerate and non-degenerate Burgers turbulence.
The reader may be interested
in direct calculation of $D'$
using \grads\ which should give $D'=D$ for
$Q_1$ and $D'=0$ for $Q_2$ distributions. As an example we present
such a derivation
for the $Q_2$-case in Section 4.

In two dimensions the distributions $Q_1$ and $Q_2$ are still different.
The interface $h({\bf x})$ is logarithmically rough (the
distribution $P(h_2,h_1,x)$ is obtained in Section 5).
We shall not make use of Loitsyanskii correlator nor we
define degeneracy above one dimension because of the lack of similarity with
NS equations.
Above two dimensions the $h$-field is bounded by geometrical reasons
and there is little difference between $Q_1$ and $Q_2$ distributions. The
bound stems from the fact that the probability of finding the interface
height $h_2$ at distance ${\bf x}$ from the given height $h_1$ is gaussian
$$P(h_2,h_1,x) \sim \exp\Big[-{\pi^{d/2}a^{2-d}(h_2-h_1)^2\over{D\Gamma(-1+d/2)
(a^{4-2d}-x^{4-2d})}}\Big],\eqn\gaussian$$
$a$ is the required latice cut-off.
The special case of a bound $H$ which is {\it smaller}
than the cut-off-related value $D^{1/2}a^{1-d/2}$
will not be considered in this paper.

\chapter{Diffusion equation. Non-bounded random initial condition.}

The solution of the diffusion
equation corresponding to \kpz\ for a given initial condition is
$$h({\bf x},t) = \int d^dy g({\bf x-y},t)h_o({\bf y}).\eqn\diff$$
For the non-bounded gaussian distribution $Q_1$ of the type \grads\ one can
calculate different correlators by using Fourier modes
$$h_o({\bf x}) = {1\over{(2\pi)^d}}\int e^{i{\bf kx}} h({\bf k})d^dk,\quad
h({\bf k})=\int e^{-i{\bf kx}}h_o({\bf x}) d^dx.$$
Solution of the diffusion equation \diff\ is given by
$$h({\bf k},t) = h_o({\bf k}) e^{-\nu{k}^2t}.\eqn\difffour$$
In terms of Fourier modes the kinetic energy density is
$$E(t) = {\lambda^2\over2}\langle ({\bf \nabla}h)^2\rangle = -
{\lambda^2\over{2(2\pi)^{2d}}}\int d^dk d^dk' kk'e^{i{\bf x}({\bf k}+{\bf k'})}
\langle h({\bf k},t)h({\bf k},t)\rangle$$
$$={\lambda^2\over{2(2\pi)^{2d}}}\int d^dk d^dk' kk'e^{i{\bf x}({\bf k}+
{\bf k'})-\nu(k^2+k'^2)t}\langle h_o({\bf k})h_o({\bf k'})\rangle.\eqn\encorr$$
The functional integral
$$\langle h_o({\bf k})h_o({\bf k'})\rangle =
\int {\cal D}[h_o]  h_o({\bf k})h_o({\bf k'}) \exp\Big[-{1\over{(2\pi)^d 4D}}
\int k^2 h_o({\bf k}))h_o(-{\bf k}) d^dk\Big].\eqn\coala
$$
equals
$$\langle h_o({\bf k})h_o({\bf k'})\rangle = 2Dk^{-2}(2\pi)^d
\delta({\bf k}+{\bf k'}),\eqn\corrnondiff$$
and gives
$$E(t) = {\lambda^2D\over{(8\pi\nu{t})^{d/2}}}.\eqn\endiffd$$

\chapter{Diffusion equation. Bounded random initial condition. One
dimension.}

This is still simple but algebraically consuming calculation which
uses mirror imaging in treating bounded gaussian
initial conditions. The representation in terms of Fourier modes is
less convenient and the calculation is performed in the original space. The
connection between kinetic energy density and correlator of the initial
field in $d$-dimensions is
$$E(t) = {\lambda^2\over2} \langle ({\bf \nabla}h)^2\rangle =
{\lambda^2\over2} \int\int d{\bf y}_1d{\bf y}_2 \big({\bf \nabla}g(y_1)
{\bf \nabla}g(y_2)\big)
\langle h_o({\bf y}_1)h_o({\bf y}_2)\rangle.\eqn\diffen$$
Integrating over half-sum $|{\bf y}_1
+ {\bf y}_2|$ and angles, we obtain the following expression
$$E(t) = {2d\lambda^2\over{\Gamma(d/2)(8\nu{t})^{1+d/2}}}\int_0^{\infty} dr
r^{d-1} e^{-r^2/8\nu{t}}\Big(1-{r^2\over{4d\nu{t}}}\Big) \langle h_o(r)
h_o(0)\rangle.\eqn\energy$$

In one dimension the weight of a given ``interface'' or rather - trajectory
which starts at some $h_o(0)=h_1$ and ends at $h_o(x)=h_2$,
$$G(h_2,h_1;x_2,x_1) = \int_{h_o(x_1)=h_1}^{h_o(x_2)=h_2} {\cal D}[h_o]
\exp\Big[-{1\over{4D}}\int_{x_1}^{x_2}\Big({dh_o\over{dy}}\Big)^2 dy
\Big],\eqn\oned$$
obeys the diffusion equation in $(h,x)$ space (not to be confused with
the original $(x,t)$ space)
$$\partial_x G = \pm D \partial_{hh} G,\eqn\hxdiff$$
here $h=h_{1,2}, x=x_{1,2}$, $\pm$ correspond to forward ($h=h_2, x=x_2
$) and backward
($h=h_1, x=x_1$) equations, respectively.
Eq\hxdiff\ is useful for imposing the constrain $|h|\leq H$ explicitly.
Namely we introduce boundary conditions $\partial_h G(\pm{H},x) = 0$ of
zero flux through the boundaries and consider the equation in the stripe
$|h|\leq H$. In order to solve the initial condition problem $G(h,h_1,0) =
\delta(h-h_1)$, we regard the $\pm{H}$ boundaries as mirrors and
sum usual Green functions of unconstrained Eq\hxdiff\ over images
$$G_{\Sigma}(h_2,h_1,x) = \sum_{m=-\infty}^{\infty} \Big[G(h_2-h_1+4Hm,x) +
G(h_2+h_1-2H+4Hm,x)\Big],\eqn\calg$$
where
$$G(h,x)=(4\pi{D}x)^{-1/2}\exp(-h^2/4Dx).\eqn\green$$

According to \energy\ one needs to evaluate the correlator $\langle h_o(x)
h_o(0)\rangle$ which is given by
$$\langle h_o(x)h_o(0)\rangle = {1\over{2H}} \int_{-H}^H
\int_{-H}^H dh_1dh_2 h_1h_2 G_{\Sigma}(h_2,h_1,x).\eqn\aaa$$
The factor $1/2H$ is due to averaging over possible values of $h_1$ which
are clearly uniform in $[-H,H]$.
Performing the elementary integration and making use of Appendix A we find
$$\langle{h_o(x)h_o(0)}\rangle = {32H^2\over{\pi^4}} \sum_{n=0}^{\infty}
{e^{-{\pi^2{D}|x|(2n+1)^2\over{4H^2}}}\over{(2n+1)^4}}.\eqn\correla$$
In the integral \energy, consider the late time asymptotic, $t\gg H^4/D^2\nu$,
then
$$E(t) = {32\lambda^2H^4\over{\pi^5D(2\pi\nu{t})^{3/2}}}.\eqn\degoned$$
Exact formula containing
a sum over
parabolic cylinder functions can be derived from \energy, \correla\ for
arbitrary times.
Summarizing, at short
times $t<<H^4/D^2\nu$, we have energy decay given by
Eq\endiffd, $E(t) \propto t^{-1/2}$, and after the crossover the decay
proceeds faster,  $E(t) \propto t^{-3/2}$.

Correlator \correla\ allows
explicit calculation of the Loitsyanskii correlator \loi.
One finds
$$\langle{v(x)v(0)}\rangle = - \lambda^2\partial_{xx}
\langle{h_o(x)h_o(0)}\rangle =
2D\lambda^2\delta(x) - {2D^2\lambda^2\over{H^2}}\sum_{n=0}^{\infty}
\e^{-{\pi^2{D}|x|(2n+1)^2\over{4H^2}}}.\eqn\corrvel$$
This sum can be rewritten by using $\theta_2$ elliptic function.
Performing the remaining integration over $x$ in \loi\ we
find that the brownian $\delta$-correlator just cancels the
non-local part generated by $|h|\leq H$ constraint, so that finally $D'=0$.
The geometrical interpretation of degeneracy leads naturally to the
potential associated with fluctuations of $h$-field. The $Q_2$ distribution
corresponds to a potential well with infinite walls at $|h|=H$.

\chapter {Green functional in two dimensions}

We continue our preliminary study of diffusion in order to develop the
method and have linear results for comparison with Burgers equation.
The derivation of the Green functional is analogous to the oscillator
problem in quantum mechanics $^{14,11}$.
We return to Eqs\pathint, \liou\
and consider a functional
$${\cal F}[h_2(\phi),h_1(\phi);y_2,y_1] = \int_{h(\phi,y_1)=h_1(\phi)}
^{h(\phi,y_2)=h_2(\phi)}{\cal D}[h] \exp\Big[-{1\over{4D}}
\int_{y_1}^{y_2}ydy \int_0^{2\pi}d\phi ({\bf
\nabla}h)^2\Big].\eqn\twod$$
We prefer to use polar system of coordinates where ``time'' is
the radius $y$.
In what follows
the adjective ``functional'' will be omitted in many cases.

The proper normalization of ${\cal F}$ will be specified later,
we need to evaluate ${\cal F}$ first.
Introducing Fourier modes for angular dependence
$$h(\phi) = \sum_{k=-\infty}^{\infty}h_ke^{ik\phi},\quad  h_k = {1\over{2\pi}}
\int_0^{2\pi}d\phi h(\phi) e^{-ik\phi},\eqn\fo$$
and changing variables to real and imaginary parts, $h_{\pm{k}} =
\alpha_k \pm i\beta_k$ we re-write \twod\ as
$${\cal F}[h_2(\phi),h_1(\phi);y_2,y_1] =
\prod_{k=0}^{\infty} I_k(\alpha_k)I_k(\beta_k),\eqn\inerm$$
where massive one-dimensional theories are defined for each mode
$$I_k(\gamma) = \int_{\gamma_1}^{\gamma_2}{\cal D}[\gamma]
\exp\Big\{-{\pi\over{
(1+\delta_{k})D}}\int_{y_1}^{y_2}dy
y\Big[\Big({\partial\gamma\over{\partial y}}\Big)^2 +
{k^2\gamma^2\over{y^2}}\Big]\Big\}.\eqn\i$$
To evaluate this gaussian integral we first determine the classical trajectory
defined by the Euler equation
$$y^2\gamma''+y\gamma'-k^2\gamma = 0.\eqn\euler$$
This equation has the solution
$$\gamma = C_1y^k + C_2y^{-k},\eqn\esolk$$
with two constants of integration to satisfy the boundary conditions
$\gamma(y_1)=\gamma_1, \gamma(y_2)=\gamma_2$.
The solution is
$$\gamma(y) = {\gamma_1y_1^k - \gamma_2y_2^k\over{y_1^{2k}-y_2^{2k}}} y^k +
{y_1^ky_2^k\over{y^k}} {\gamma_2y_1^k - \gamma_1y_2^k\over{y_1^{2k}-y_2^{2k}}}.
\eqn\solk$$
The zeroth mode, $k=0$ has a different trajectory,
$$\gamma(y) = {\gamma_1\ln{y_2}- \gamma_2\ln{y_1}\over{\ln{y_2\over{y_1}}}} +
{\gamma_2 - \gamma_1\over{\ln{y_2\over{y_1}}}} \ln{y}.\eqn\solo$$
Calculating the classic action we get
$$ -{{\pi}
k\over{D}}[(\gamma_1^2+\gamma_2^2) f_s -2\gamma_1\gamma_2
f_c]\eqn\ac$$
for non-zero $k$ and
$$ -{\pi\over{2D}}{(\gamma_2-\gamma_1)^2\over{\ln{y_2\over{y_1}}}}
\eqn\aco$$
for the zeroth mode. Here the functions of stereographic projection are
introduced
$$f_s = (r^2+1)/(r^2-1),\quad f_c = 2r/(r^2-1),\quad r=(y_2/y_1)^k,\eqn\fcs$$
and
$$f_s^2-f_c^2 = 1.\eqn\iden$$

Returning to the
functional one gets
$${\cal F}[h_2,h_1;y_2,y_1] = C \exp\Big[-
{\pi(\alpha_{2,0}-\alpha_{1,0}
)^2\over{2D\ln(y_2/y_1)}}\Big]\times$$
$$ \exp\Big\{ - \sum_{k=1}^{k=\infty}
{\pi{k}\over{D}} \Big[\Big(\alpha_{1,k}^2+
\beta_{1,k}^2+\alpha_{2,k}^2+\beta_{2,k}\Big) f_s
- 2\Big(\alpha_{1,k}\alpha_{2,k}+
\beta_{1,k}\beta_{2,k}\Big) f_c\Big]\Big\},\eqn\ph$$
or, in terms of the original Fourier modes,
$${\cal F}[h_2,h_1;y_2,y_1] = C \times $$
$$ \exp\Big\{ - \sum_{k=-\infty}^{\infty}
{\pi{k}\over{D}} \Big[\big(h_{1,k}h_{1,-k}
+h_{2,k}h_{2,-k}\big)f_s -
\big(h_{1,k}h_{2,-k} + h_{1,-k}h_{2,k}\big)f_c\Big]\Big\}.\eqn\pha$$
Note that the factors $kf_{s,c}$ are even in $k$ and have proper limits, so
that the summation in \pha\ is
expanded to all integers, including $k=0$.

Integrating ${\cal F}$ over (say) final field $h_2
$, one finds that it is not
normalized. The reason is that the representation \inerm, \i\ descibes a
massive
theory, and classic action becomes zero only if all $h_{1,k} = 0$ for $k\neq
0$.
The zeroth
mode can be independently normalized in Eq\pha. Other modes, if normalized
forcibly, give rize to inconvenient non-local factors which do not ensure
convolution properties.
Thus,
one has to perform calculation with the functional
${\cal F}$ and normalize the answer for all non-zero modes
with respect to bare theory at the very end of calcualtion.
With this in mind ${\cal F}$ can be considered as the {\it Green functional}.

We then
derive the differential
equation to which the integral \twod, \pha\ satisfies.
In complete analogy with the one-dimensional case $^{14}$
we consider small increment  $y_2+\delta{y}$ and the corresponding variation of
the field $h_2$:
$${\cal F}[h_2+\eta,h_1;y_2+\delta{y},y_1] =
\int {\cal D} \eta {\cal F}[h_2+\eta,h_1;y_2,y_1]
\times$$
$$\exp\Big\{-{y_2\over{4D \delta{y}}} \int_0^{2\pi}d\phi \eta^2 -
{ \delta{y}\over{4D y_2^2}} \int_0^{2\pi}d\phi \Big({dh_2\over{d\phi}}\Big)^2
\Big\}.\eqn\vari$$
Expanding ${\cal F}[h_2+\eta,h_1;y_2,y_1]$ to the second order in variation
$\eta$ and integrating out $\eta$
one finds that the linear term in $\eta$ vanishes
and the second order derivative is non-zero only if the arguments of
$h_2(\phi)$
functions are equal. Expanding also the $\delta{y}$-dependence to the first
order, we finally obtain
$$\partial_y{\cal F} = {D\over{y}}\int_0^{2\pi} d\phi {\delta^2{\cal F}\over{
\delta{h}^2}} - {{\cal F}\over{4Dy}}
\int_0^{2\pi} d\phi \Big({dh\over{d\phi}}\Big)^2,\eqn\fdiff$$
where subscript 2 can be omitted. We shall use the symbolic notation for
the diffusion operator of closed strings and write Eq\fdiff\ as
$$\partial_y{\cal F} = \hat{D}{\cal F}.\eqn\ha$$
Clearly, the normalization condition is not satisfied: Eq\fdiff\
contains a decay term. The solution of Eq\fdiff\ is
$${\cal F}[h_2,h_1;y_2,y_1] = C' \prod_k \Big({k f_c\over{2D}}
\Big)^{1/2} \times
$$
$$\exp\Big\{ - {\pi{k}\over{2D}}
\Big[\big(h_{1,k}h_{1,-k}
+h_{2,k}h_{2,-k}\big)f_s -
\big(h_{1,k}h_{2,-k} + h_{1,-k}h_{2,k}\big)f_c\Big]\Big\},\eqn\phorig$$
where $C'$ is now independent upon $h_1, h_2, y_1, y_2$.
This formula contains the explicit prefactor of the functional ${\cal F}$\
\pha\ and was found by T.J. Newman.
 The prefactor can be alternatively found by computing
gaussian fluctuations around the classic trajectory.
With the help of the prefactor
the convolution property can be proved
$${\cal F}[h_2,h_1;y2,y1] = \int {\cal D}[h_3] {\cal F}[h_2,h_3;y_2,y_3]
{\cal F}[h_3,h_1;y_3,y_1].\eqn\convol$$
The following identities are useful
$$f_c(y_3/y_1)f_c(y_2/y_3)[f_s(y_3/y_1)+f_s(y_2/y_3)] = f_c(y_2/y_1),$$
$$f_s^2(y_3/y_1)-f_s^2(y_2/y_3) = f_c^2(y_2/y_1).\eqn\identt$$

The Green functional \phorig\ is connecting angular $h(\phi)$-profiles
between two non-zero radial ``times'' $y_1$ and $y_2$, $y_2>y_1$. In some
cases we shall need to set $y_1=0$. The logarithmic dependence in \phorig\
leads to a divergence here.
Divergence indicates that the weight $({\bf\nabla}h)^2$
in the action is insufficient to ensure convergence of the functional.
This problem is usually resolved by introducing UV cut-off,
$a$. In a proper
field theory it is anticipated that the cut-off present in
bare values must be eliminated
by renormalization when interaction is included. However, there is
{\it no} reason to expect that the Burgers turbulence is a
proper field theory.

In some cases we shall need the angular part of the Green functional
\phorig\ to be integrated out. This happens, for example, when
one calculates the two-point correlator, $\langle{h(x)h(0)}\rangle$.
Although this correlator does not formally exist in the unbounded case,
it is useful to define the probability $P(h_2,h_1,x)$
of arriving to $h(x)=h_2$ provided that $h(0)=h_1$. It is given by the
integral (to be normalized)
$$P(h_2,h_1,x) \sim \int {\cal D}[h_2(\phi)]{\cal F}[h_2(\phi),h_1;x,a]
\delta(h_2(\phi_0)-h_2).\eqn\ppp$$
It is understood in Eq\ppp\ that the the two points needed to define
$P(h_2,h_1,x)$ are the origin and $(x,\phi_0)$ in polar coordinates of
\phorig. The angle $\phi_0$ is arbitrary. Exponentiating $\delta$-function
and performing gaussian integration one finds (c.f. \gaussian)
the two-point probability
$$P(h_2,h_1,x) = {1\over{\sqrt{4D\ln(x/a)}}}
\exp\Big[-{\pi(h_2-h_1)^2\over{4D\ln(x/a)}}\Big],\eqn\probb$$
with the obvious normalizing prefactor.
This function obeys a diffusion-like PDE,
$$\partial_x P = {D\over{\pi{x}}}\partial_{hh}P,\eqn\ze$$
with a cut-off at small $x$, $x\geq a$. Eq\ze\ describes a logarithmically
wandering interface $^{15}$.

\chapter {Diffusion equation. Bounded random initial condition.
Two dimensions}

In this Section we derive the exponents for the kinetic energy decay in
two dimensions and obtain the behavior resembling $O(2)$ non-linear
$\sigma$-model after the onset
of bounded properties in the initial condition.
The method of mirror images is again useful.
Due to translational invariance and
isotropy of the problem the only correlator that
we need is $\langle h(x)h(0)\rangle$.
The probability $P(h_2,h_1,x)$ given by Eq\probb\
is sufficient to perform the calculation if used instead of $G(h,x)$, Eq\green.
Function $P$ can be obtained from $G$ by replacements $D\rightarrow D/\pi$ and
$x\rightarrow \ln{x\over{a}}$. The correlator \correla\ takes the form
$$\langle{h_o(x)h_o(0)}\rangle = {32H^2\over{\pi^4}} \sum_{n=0}^{\infty}
{1\over{(2n+1)^4}}\Big({x\over{a}}\Big)
^{-{\pi{D}(2n+1)^2\over{4H^2}}}.\eqn\correlb$$
Integration in \energy\ with the low limit being $x=a$ yields
$$E(t) = {8\lambda^2
H^2\over{\pi^4\nu{t}}} g\Big({8\nu{t}\over{a^2}}\Big),\eqn\entwo$$
with function
$$g(x) = \sum_{m=0}^{\infty}{x^{-s(m)}\over
{(2m+1)^4}} [\Gamma(1-s(m),
x^{-1})-\Gamma(2-s(m),
x^{-1})],
\eqn\innn$$
and $s(m)={\pi}D(2m+1)^2/8H^2$. $\Gamma(a,b)$ is the incomplete
gamma-function.
The limit $\nu{t}\gg a^2$ appears to be rich and is studied in
Appendix B. Following Appendix consider the cases:

1). $D/H^2<8/\pi$. Using (B.1) we get
$$E(t) = {\lambda^2
D\over{\pi^3\nu{t}}}\Big({8\nu{t}\over{a^2}}\Big)^{-\pi{D}/8H^2}
\Gamma\Big(1-{\pi{D}\over{8H^2}}\Big) \propto t^{-1-\pi{D}/8H^2}.\eqn\les$$
The decay exponent depends linearly upon the parameter
$D/H^2$ and changes from $-1$ (c.f. non-bounded case) down to $-2$ at
the critical value $D/H^2 = 8/\pi$. Note that the corrections to Eq\les\
are of the order of $t^{-1-s(m)}$ so that there is an accumulation
point at $D/H^2=0$. To get the prefactor in the limit $D/H^2\rightarrow0$
(which is, of course, given by \endiffd)
it is easier to return to
the integral \energy. The following formula is helpful
$$\int_0^{\infty}d\xi (\xi-1)e^{-\xi} \ln\xi  = 1.$$

2). $D/H^2 = 8/\pi$. There is a logarithmic correction in the
leading order,
$$E(t) = \Big({\lambda
{aH}\over{\pi^2\nu{t}}}\Big)^2\ln\Big({8\nu{t}\over{a^2}}\Big).\eqn\oneexx$$

3). $D/H^2 > 8/\pi$.
Using (B.3), (B.4) one finds
$$E(t) = {1\over{4\pi^3}} \Big({\lambda{aH^2}\over{\nu{t}}}\Big)^2
\Big[s(0)\tan\Big({\pi\over{2\sqrt{s(0)}}}\Big)-{\pi{s(0)}\over{2}}-{\pi^3\over{
24}}\Big] \propto t^{-2}.\eqn\dal$$

We note that the dependence of the decay exponent on the parameter
$D/H^2$ is similar to the $O(2)$ non-linear $\sigma$-model $^{10,11}$.
This similarity is not complete, however. Recall that the spin direction, which
distribution is described by $Q_{2}$, is a {\it cyclic} variable, like
our mirror imaged field $h_o$.
At non-zero temperature (in our case - at non-zero $D$) there exist
vortex excitations which interact like Coulomb gas. For our field $h_o$
this would mean the presence of point defects in the vicinity of which
$h_o(y,\phi) = 2H\phi$. The velocity field ${\bf v}(y,\phi) =
-(\lambda/y){\bf e}_{\phi}$ is not curl-free. Recall now that
the absence of vorticity was assumed by Burgers
when he ``derived'' his equation from NS equation. To make consistent
comparison with Burgers equation
we avoid vorticity at all stages. Other applications such as
nonlinear heat flow and interface growth also do not lead to
point defects under usual circumstances.
Consequently, our correlators behave as
low-temperature correlators in the $\sigma$-model $^{10,11}$. To show this,
one may use the probability $P(h_2,h_1,x)$ to calculate the average
$\langle\cos(h_2-h_1)\rangle$ usually considered in spin models. It is given
by
$$\langle\cos(h_2-h_1)\rangle = {1\over2}\Big({x\over{a}}\Big)
^{-{\pi{}D\over{H^2}}} +
\sum_{n=0}^{\infty} {8(2n+1)^2\over{\pi^2[(2n+1)^2-4]^2}}
\Big({x\over{a}}\Big)^{-{\pi{}D(2n+1)^2\over{4H^2}}}$$
$$
\rightarrow {8\over{9\pi^2}}
\Big({x\over{a}}\Big)^{-{\pi{}D\over{4H^2}}}
.\eqn\nons$$
This correlator can not diminish steeper than $x^{-2}$ (see Ref. 11) and
therefore $D/H^2 = 8/\pi$ is the largest value of $D/H^2$ when \nons\
(and distribution $Q_2$) represent the spin model. This is precisely the
transition value of $D/H^2$ found above. It is eight times larger than the
Kostelitz-Thouless transition temperature $^{11}$
(to indicate the connection, one
assumes $k_BT/J = 2D$, $H = \pi/2$, so that $k_BT_c/J = \pi/2$ corresponds to
$D_c/H^2 = 1/\pi$).

In this respect it may seem surprizing to find
the abrupt transition at $D^2/H^2 = 8/\pi$. The analog of steepest correlator
\nons\ in our study is
the kinetic energy density of the {\it discontinous}
bounded distribution which diffusively decays as $t^{-2}$. The relaxation of
continuous fields should not be faster. Thus, kinetic energy is a curious
quantity which undergoes transition with mirror images in the absence of
topological charges.
The transition is a special property of the sharp features of the
potential associated with $Q_2$-distribution. To clarify
this one may calculate the kinetic energy decay for the gaussian
distribution
$$Q_3 \sim \exp\Big[-{1\over{4D}}
\int d^dx[(\nabla{h})^2 + m^2h^2]\Big],\eqn\qthree$$
and find
that there is a single crossover time at $t = 1/\nu{m^2}$, which is
$D$- and $d$-independent. The energy is given by
$$E(t) = {\lambda^2 D d m^d\over{2(4\pi)^{d/2}}}
\e^{2m^2\nu{t}}\Gamma\Big(-{d\over{2}},2m^2\nu{t}\Big).\eqn\massi$$
At short times
one observes the non-bounded decay \endiffd\ and after the crossover the
decay reaches its maximum rate $t^{-d/2-1}$ without any unusual behavior
in two dimensions. Thus, the onset of bounded behavior is sensitive to
the shape of the potential which provides the boundary: the parabolic
potential of the distribution \qthree\ leads to a behavior different
from that of infinite well potential $Q_2$.

Returning to vortices, we note that diffusion equation and
(formally) Burgers equation may be
considered with random initial conditions having solenoidal component.
We are not aware of possible physical applications, though.
It is interesting to mention that for both equations
the decay of vorticity is pure
diffusional, while the potential component may have rich behavior
since it is coupled to the solenoidal component.

\chapter{Burgers turbulence. Bounded random initial distribution. One
dimension.}

We have completed our preliminary steps of deriving the reference results
for diffusion equation and begin to study Burgers equation in the form
of the field theory reproduced in Section 2.
The case of non-bounded initial condition has been considered in Ref.1.
We denote as $\Phi(h,y)$ the
density of the function $\psi$ (1.5) along the $h$-axis; it is a sum over all
the paths terminating at the point $h$ for a given ``time'' $y$.
This
function obeys the forward and backward
Shrodinger equation $^1$(diffusion equation with a decay term)
$$\partial\Phi(h,y)/\partial{y} =
\pm D\partial\Phi/\partial{h^2} \mp u g(y)
e^{\lambda h\over{2\nu}}\Phi, \eqn\main$$
which is considered in the strip $|h|\leq H$ (c.f. Section 4).
The boundary conditions are $\partial_h \Phi(\pm{H},y) = 0$, and the
``initial''
condition reflects uniform distribution in the strip far from the origin $y=0$,
i.e. far from the potential support (decay support): $\Phi(h,\mp\infty) =
1/2H$.

Eq\main\ can be simplified in the limit of zero viscosity, $\nu\rightarrow 0$.
Introducing new variable $h_u = - (2\nu/\lambda) \ln u$ one can rewrite
\heig\ as
$$\langle{h}\rangle = \int_{-\infty}^{\infty} dh_u \Big[ \exp(-e^{-{\lambda h_u
\over{2\nu}}}) - \psi(h_u)\Big] \rightarrow \int_{-\infty}^{\infty} dh_u \Big[
\Theta(h_u) - \psi(h_u)\Big],\eqn\heighu$$
where $\Theta(z)$ is the step-function. The decay term in Eq\main\ becomes
a $\delta$ - function, i.e. there appears an absorbing curve
$$h = h_u + y^2/2\lambda{t};\eqn\lin$$
and pure diffusion equation
$$\partial\Phi(h,y)/\partial{y} =
\pm D\partial\Phi/\partial{h^2} \eqn\maindiff$$
below this parabolic curve. We can see now that the
Burgers approach $^2$ is equivalent to the present study in the limit
$\nu\rightarrow 0$.

For some time there is no difference between bounded and unbounded cases since
local properties of the distributions $Q_1$ and $Q_2$ are almost identical.
In the limit of strong turbulence that we are interested in, $t\gg \nu^3/D^2
\lambda^4$, the energy decays as $^{2,1}$
$$ E = \eta \Big({D\lambda^2\over{t}}\Big)^{2/3},\eqn\initi$$
where $\eta$ is a number $^{16}$. The averaged height of the corresponding
interface problem grows as $\langle{h}\rangle = 3\eta (D^2\lambda{t})^{1/3}$.
This regime ends when the averaged height approaches the limiting value $H$,
since we have $\langle{h}\rangle < H$ by the above-mentioned boundary
conditions.
At later times the parabola \heighu\ is quite close to the upper boundary $H$.
One can then approximate the parabola by making the boundary $h = H$ absorbing
within the region $|y| < y_0$, where
$$y_0 = \sqrt{2\lambda{t}(H-h_u)},\eqn\cf$$
see \heighu. Selecting the forward version of the diffusion equation \maindiff,
we start with the function $\Phi(h)$ uniform in $[-H,H]$ at the
point $y=-y_0$. The solution at the end of the absorbing region, $y=+y_0$ is
given by
$$\Phi(h,y) = \int_{-H}^H dh' {1\over{\sqrt{8\pi{D}y_0}}}
\sum_{m=-\infty}^{\infty}
\Big[e^{-{(h-h'+4Hm)^2\over{8Dy_0}}} - e^{-{(h+h'-2H+4Hm)^2\over{8Dy_0
}}}\Big].
\eqn\ppph$$
The minus sign here (c.f. \calg) accounts for absorbing boundary. It follows
from \heighu\ that to find the
averaged height $\langle{h}\rangle$ one has to evaluate the integral
$$\langle{h}\rangle = \int_0^H dh_u \Big[1 - \int_{-H}^H dh \Phi(h,y)\Big] =
H - \int_0^H\int_{-H}^H dh_udh\Phi(h,y).\eqn\ergh$$
Using another representation (A.1)
of the elliptic functions entering \ppph\
one obtains that $H-\langle{h}\rangle$ equals
$$\int_0^H\int_{-H}^H\int_{-H}^H
dh_udhdh'
\sum_{n=1}^{\infty} e^{-{\pi^2n^2Dy_0\over{2H^2}}} \Big\{\cos
\Big[{\pi{n}\over{2H}}
(h-h')\Big] - \cos\Big[{\pi{n}\over{2H}}(h+h'-2H)\Big]\Big\}$$
$$ = {8\over{\pi^2}}
\int_0^Hdh_u \sum_{l=0}^{\infty}
{e^{-{\pi^2Dy_0(2l+1)^2\over{2H^2}}}\over{(2l+1)^2}}.\eqn\wrt$$
With the help of \cf\
one gets the late time asymptotic behavior for the
averaged height
$$\langle{h}\rangle = H - {H^4\over{30 D^2 \lambda{t}}},\eqn\trans$$
and the kinetic energy decay
$$E = {H^4\over{30 D^2 t^2}}.\eqn\entrans$$

This is a very fast decay for one dimensional problem,
faster then the asymptotic behavior of pure diffusion
\degoned. One then expects a second crossover to pure diffusion
if finite viscosity is allowed back in
Eq\main. The diffusive behavior at late times must be accessible by
direct perturbation in the decay strenght.
Let us introduce an auxilary function
$$\Psi(y) = \int_{-H}^{H}\Phi(h,y)dh,\quad \Psi(\infty) = \psi.\eqn\psidef$$
To zeroth order in $u$ we have $\Phi_0(h,y) = 1/2H$, this is essentially an
adiabatic approximation. Decay influences the {\it amplitude} of the solution,
to the first order
$$\Phi(h,y) = \Psi_1(y)/2H.\eqn\psione$$
Function \psione\ upon substitution into Eq\main\ and integrating over $[-H,H]$
has the form
$$\Psi_1(y) = \exp\Big[-u{\sinh\kappa\over{\kappa}}\int_{-\infty}^ydy
g(y)\Big],
\eqn\psioneone$$
where the Reynolds number is defined as $\kappa = \lambda{H}/2\nu$.
As usual the knowledge of solution to zeroth order  is enough to calculate
the leading behavior of the integrated properties. Returning to the averaged
field $h$, which is given by Eqs\heig,
\psidef\ and, to this order, by Eq\psioneone, we
get
$$\langle h\rangle = {2\nu\over{\lambda}} \ln\Big({\sinh\kappa\over{\kappa}}
\Big).\eqn\satur$$
In terms of KPZ equation \kpz\ this
implies that the averaged height ceases to move at late times and saturates
at the level \satur\, below $H$. We again expect from this result
that the final stage of decay is pure diffusion, since saturation means
small gradients and irrelevance of $\lambda$-term in \kpz.

One has to determine the correct
amplitude of the diffusion-like decay, $E(t) \propto t^{-3/2}$ (see Section 4
for the derivation of this power-law).
To go beyond zeroth order in $u$ regarding Eq\main\ we make use of the
expansion
$$\Phi(h,y) = {\Psi_1(y)\over{2H}}
[1+u\mu_1(h,y)+{u^2\over2}\mu_2(h,y)+...],\eqn\expan$$
which is inspired by WKB approximation.
Substituting \expan\ into Eq\main\ and collecting similar terms we get the
equation for the first-order in $u$ corrections to the solution
$$\partial_y\mu_1 - D\partial_{hh}\mu_1 = g(y)\Big({\sinh\kappa\over{\kappa}}
- e^{\lambda{h}\over{2\nu}}\Big).\eqn\firstor$$
Its solution is
$$\mu_1(h,y) = \int_{-\infty}^{y}dy'\int_{-H}^Hdh' g(y')
\Big({\sinh\kappa\over{\kappa}}- e^{\lambda{h'}\over{2\nu}}\Big){\cal G}
(h,h',y-y'),\eqn\solfirst$$
where the appropriate Green function ${\cal G}$ is defined by \calg.
It is easy to check that
$$\int_{-H}^H dh \mu_1(h,y) = 0,\eqn\che$$
so that there is no contribution to function $\psi$ in this order.
Second order in $u$ gives the equation
$$\partial_y\mu_2 - D\partial_{hh}\mu_2 = g(y)\mu_1\Big(
{\sinh\kappa\over{\kappa}}- e^{\lambda{h}\over{2\nu}}\Big).\eqn\secondor$$
Integrating this equation with the help of ${\cal G}$ we may represent the
quantity of interest, $I = \int_{-H}^H dh \mu_2(h,\infty)$ in the form
$$I = \int^{\infty}_{\infty}\int_{\infty}^{y_2} dy_2 dy_1
g(y_1,t)g(y_2,t)\int_{-H}^H \int_{-H}^H dh_1 dh_2 Z(h_1)Z(h_2) {\cal
G}(h_2,h_1,y_2-y_1), \eqn\call$$
where
$$Z(h)=\sinh(\kappa)/\kappa-\exp({\lambda}h/2\nu).\eqn\ff$$
The evaluation of the integral \call\ follows the same lines as \aaa.
We first change variables $y=y_2-y_1, z=y_2+y_1$ and integrate out $z$. One
finds
$$I = {1\over{\pi\sqrt{8D{\nu}t}}}\int_0^{\infty}
{dy\over{\sqrt{y}}}e^{-y^2/8{\nu}t} J(y),\eqn\jj$$
where
$$J(y) = \int_{-H}^H \int_{-H}^H dh_1 dh_2 Z(h_1)Z(h_2)
\sum_{m=-\infty}^{\infty} \Big[e^{-{(h_2-h_1+4Hm)^2\over{4Dy}}} +
e^{-{(h_2+h_1-2H+4Hm)^2\over{4Dy}}}\Big].\eqn\jjj$$
Changing variables and using symmetry properties of the integrand we obtain
$$J(y) = 4H^2 \int_0^1 d\eta p(\eta)
\sum_{m=-\infty}^{\infty} e^{-{H^2\over{Dy}}(\eta+2m)^2},\eqn\jjjj$$
with
$$p(\eta) = \int_{-1+\eta}^{1-\eta}d\eta' Z[H(\eta'-\eta)]Z[H(\eta'+\eta)] +$$
$$\int_0^{\eta}d\eta'\{{1\over2}Z[H(\eta-\eta'-1)]Z[H(\eta+\eta'-1)] +
{1\over2}
Z[H(1-\eta-\eta')]Z[H(1-\eta+\eta')]\} = $$
$$= {1\over4}\kappa^{-2} e^{-2\kappa(1+\eta)}\big(e^{2\kappa}-e^{4\kappa}-
e^{2\kappa\eta}-e^{4\kappa\eta}+2e^{2\kappa(1+\eta)}-e^{2\kappa(2+\eta)}+
$$
$$e^{2\kappa(1+2\eta)}+2\kappa{e^{4\kappa}}-2\kappa
{e^{4\kappa\eta}}
-\eta{e^{2\kappa\eta}}+2\eta{e^{2\kappa(1+\eta)}}-\eta{e^{2\kappa(2+\eta)}} +
2\kappa^2\eta{e^{4\kappa}}+2\kappa^2\eta{e^{4\kappa\eta}}\big).\eqn\ppp$$
Note, that $\int_0^1d\eta{p(\eta)} = 0$. Now, the sum entering \jjjj\ can be
re-written using (A.1). Denoting $p_l = \int_0^1d\eta p(\eta)\cos(\pi{l}\eta)$,
and integrating over $y$ we find
$$I = 4H \sum_{l=1}^{\infty} p_l
\exp\Big[2\nu{t}\Big({\pi^2l^2D\over{4H^2}}\Big)^2\Big]
{\rm erfc}\Big[\sqrt{2\nu{t}}{\pi^2l^2D\over{4H^2}}\Big].\eqn\iiiii$$
The applied expansion \expan\ is consistent when $D$-diffusion
in $h$ direction
is faster then $\nu$-diffusion in $y$ direction, i.e. $D^2\nu{t}/H^4 \gg 1$.
Using the asymptotic form of the error-function  and returning to kinetic
energy, we get
$$E(t) = 2\nu \partial_t \int_0^{\infty}{du\over{u}}\Big[e^{-u} - \Psi_0 -
{u^2\over{2}}{1\over{2H}}\Psi_0 I\Big] =
{4\nu{H}^2\over{\pi^2\sqrt{2\pi\nu{t}}Dt}}
{\kappa^2\over{\sinh^2\kappa}} \sum_{l=1}^{\infty}{p_l\over{l^2}}.\eqn\eee$$
The remaining sum can be evaluated by using the definition of $p_l$ and \ppp\.
$$\sum_{l=1}^{\infty}{p_l\over{l^2}} =
{\pi^2\over{192\kappa^4e^{2\kappa}}}\big(
21-42e^{2\kappa}+21e^{4\kappa}+30\kappa-30\kappa{e^{4\kappa}}+13\kappa^2+
10\kappa^2e^{2\kappa}+13\kappa^2e^{4\kappa}\big).\eqn\fisum$$
The asymptotical form in the limit of strong turbulence $\kappa\gg 1$ is
$$E(t) = {13\nu^2H^2\over{48\sqrt{2\pi}D(\nu{t})^{3/2}}},
\eqn\ajs$$
a diffusive decay which differs from \degoned\ only by its amplitude.

We may now reconstruct the entire decay process. At short times, when
degeneracy
of the turbulence is not important
the decay is given by the Burgers formula
\initi.
The averaged height $\langle{h}\rangle$ approaches $H$
at $t_1 = H^3/D^2\lambda$.
{}From the other side the diffusive decay \ajs\ applies at $D^2\nu{t}/H^4 \gg
1$,
when the $\nu$-diffusion lenght $(\nu{t})^{1/2}$ exceeds $H^2/D$.
This defines $t_2 = H^4/D^2\nu$, the second crossover time. The ratio of the
two times is $t_2/t_1 \sim \kappa$.
In the case of strong turbulence, $\kappa\gg1$, these
results indicate the existence of an intermediate regime at $t_1\ll t\ll t_2$
(if $\kappa\ll 1$, the entire process is just diffusion). The intermediate
regime is described by \entrans.
We have made numerical simulations of Eqs\main, \maindiff. The results are
shown in Figs.1-4.

\chapter{Burgers turbulence in
two dimensions}

Using the results of Section 5 we introduce the density of the functional
integral \pathint\ and by
analogy with the previous Section write the functional diffusion
(Schrodinger) equation to which this density satisfies
$$\partial_y\Phi[h(\phi),y] = \pm \hat{D}
\Phi \mp uyg(y) \Phi \int_0^{2\pi} d\phi e^{\lambda h(\phi)\over{2\nu}}.
\eqn\maint$$
The equation is considered on the space of all $h(\phi)$ profiles
and
radius-``time'' $y$ exceeds the cut-off, $y>a$.
The derivation of the operator $\hat{D}$ implies that
$\Phi$ is a double density, with specified initial and final profiles.

We shall only consider the invisid limit for the non-bounded case $Q_1$.
Even in the limit of vanishing viscosity $\nu$ the ratio $\nu{t}/a^2$ should
be assumed large. The structure of Eq\maint\ suggests the change of
the ``time''-variable $z=\ln(y/\sqrt{4\nu{t}})$. Equation can be re-written as
$$\partial_z\Phi = D\int_0^{2\pi} d\phi {\delta^2\Phi\over{
\delta{h}^2}} - {\Phi\over{4D}}
\int_0^{2\pi} d\phi \Big({dh\over{d\phi}}\Big)^2 -
2 \Phi \int_0^{2\pi} d\phi \exp\Big[2z-e^{2z}+
{\lambda [h(\phi)-h_u]\over{2\nu}}\Big],\eqn\hu$$
where we selected the forward version and introduced the familiar
parameter $h_u = - (2\nu/\lambda)\ln{u}$ (see Section 7). Eq\hu\ exibits
two decay terms. One of them comes from the diffusion of strings and the
other is generated by the interaction term in \liou. By analogy to the
previous Section, we define a surface
$$h(\phi,z) = h_u + {2\nu\over{\lambda}}(e^{2z}-2z),\eqn\curve$$
above which the interaction-generated decay dominates.
When $\nu\rightarrow 0$ the surface \curve\ becomes a plane parallel to
$\{y,\phi\}$ plane up to the region of
large positive $z$
when the exponential term in \curve\ can no longer be neglected. Let us
denote these values of $z$ as $\bar{z}$. Below the absorbing plane $h = h_u$
we have pure diffusion equation \ha\
with the point source at
$(h,z) = (0,z_{\min})$, $z_{\min}=\ln(a/\sqrt{\nu{t}})$.
The abrorbing surface can be accounted by mirror imaging of this source
with opposite sign. Thus, the solution for $\Phi$ reads
$$\Phi[h(\phi),z] = {\cal F}\Big[h(\phi),0;{e^{\bar{z}}\over{4\nu{t}}},a\Big] -
{
\cal F}\Big[h(\phi),2h_u;{e^{\bar{z}}\over{4\nu{t}}},0\Big]\eqn\differ$$
with functional \phorig. This solution has to be integrated over all final
profiles $h(\phi)$ (or sections of these) which end below the absorbing
surface \curve. It can be conveniently done in two steps. First, we integrate
over all profiles which pass through a specified point $(\bar{z},\phi_0)$,
this leads to the two-point probability $P$, introduced in Section 5.
With the help of \heighu\ we perform the second step
$$\langle{h}\rangle = \int_0^{\infty} dh_u \Big\{ 1 - \int_{-\infty}^{h_u} dh
\Big[P\Big(h,0,{e^{\bar{z}-z_{\min}}\over{4\nu{t}}}\Big) -
P\Big(h,2h_u,{e^{\bar{z}-z_{\min}}\over{4\nu{t}}}\Big)
\Big]\Big\} =$$
$$ \int_0^{\infty} dh_u {\rm erfc}\Big[{\pi^{1/2}
h_u\over{4D(\bar{z}
-z_{\min})}}\Big]  = {1\over{\pi}}\sqrt{4D(\bar{z}-z_{\min})}.\eqn\er$$
{}From the definition of $\bar{z}$ it is clear that its time-dependence is
of the type $\bar{z} = {1\over2}\ln(\lambda\langle{h}\rangle/2\nu) \sim
\ln \ln t$. With logarithmic precision one can set $\bar{z} =0$ and
obtain
$$ \langle{h}\rangle = \sqrt{{2D\over{\pi^2}} \ln\Big({\nu{t}\over{a^2}}\Big)},
\eqn\hetwo$$
$$ E = {\lambda{D}^{1/2}\over{\pi{t}\sqrt{2\ln(\nu{t}/a^2)}}}.\eqn\ee$$
Analogous time-dependence
$t^{-1}\ln^{-1/2}t$ is obtained in Ref.4
for a  different
problem.

We briefly discuss the bounded case. As in one dimension there is a
crossover time which can be found by equating
$\langle{h}\rangle = H$, see Eq\hetwo, $t_1 = (a^2/\nu) e^{\pi^2H^2/2D}$.
In the invisid limit this is the end of the evolution.  If $\nu$ is finite,
kinetic energy quickly falls down and later is of order $(a/t)^2$.

\chapter {Conclusion}

We have presented a detailed study of Burgers turbulence decay in one and
two dimensions and we think that the decay of kinetic energy is more or less
understood.
In one dimension our realization of the degenerate turbulence indicates the
existence of three-stage relaxation which serves as a counter-example for
the prediction of the similarity hypothesis based upon zero Loitsyanskii
correlator. Obviously, zero correlator
does not fully define the
distribution
and the existence of other counter-examples
is expected.
The results is $d=2$ demonstrate
an interesting behavior
when the energy decay contains
logarithmic factors and explicit cut-off-dependence. Therefore, the
scaling (if it exists) is preceded by an exponentially long transient
requred to ensure $\ln({\nu}t/a^2) \gg 1$. The existence of exponentially
slow crossover has been recognized earlier in the noise-driven case $^{17}$.
The crossover
complicates the physics of the noise-driven case, where exponents
obtained by numerical simulations are found to be model-dependent.
We hope that this paper will stimulate numerical and
RG studies of the initial condition problems.
Subsequently, the exact results
from this method, computer simulations,
RG analysis and, hopefully, the application of the
direct integration approximation
by Kraichnan will be instructively compared at some
stage.

\ack{It is a pleasure to thank T.J. Newman
with whom the first part (Ref. 1) of this
study was written and many of the above given results were
actively discussed. I am grateful to G.L. Eyink, D. Frenkel
and
N. Goldenfeld for discussions and to J. Krug for communicating the
review, Ref. 12.
This work was supported in part by the Material Research Laboratories at the
University of Illinois at Urbana-Champaign and at the University of Chicago,
and in part by NSF Grant NSF-DMR-90-15791.

\chapter{Appendix A. Two expansions of $\theta_3$ elliptic function}

The following equality is useful in treating mirror images
$$\sum_{m=-\infty}^{\infty} e^{-(\eta+2m
)^2/x} = {\sqrt{\pi{x}}\over2}\theta_3\Big({\pi\eta\over2},
e^{-{\pi^2x\over4}}\Big) = {\sqrt{\pi{x}}\over2}\Big[1+2
\sum_{m=1}^{\infty} e^{-{\pi^2m^2x\over4}}\cos(\pi{m}\eta)\Big].$$

\chapter{Appendix B. Function $g(x)$ defined by \innn.}

Consider the limit $x \gg 1$.
There is a special sequence $s(m)=1,2,3...$ when logarithmic terms appear
in the expansion of incomplete gamma-function and different
corrections to the main dependence acquire logarithms.
One can identify three different situations.

1). $s(0) < 1$.  We find
$$g(x)=s(0)x^{-s(0)}\Gamma(1-s(0)).\eqn(B.1)$$

2). $s(0) = 1$. There is a logarithmic correction in the
leading order,
$$g(x) = {\ln x\over{x}} - \Big(\gamma + {\pi^4\over{96}}\Big) + O(x^{-2}).\eqn
(B.2)$$

3). $s(0) > 1$.
The leading dependence can be
shown to be
$$g(x) = {1\over{x}}\Big[{\pi\over{4}}s(0)\tan\Big({\pi\over{2\sqrt{s(0)}}}
\Big) - {\pi^2s(0)\over8} - {\pi^4\over{96}}\Big] + O(x^{-s(0)}) + O(x^{-2}).
\eqn(B.3)$$
with the asymptotics
$$xg(x) = {1\over{s(0)-1}} - {\pi^4+12\pi^2-216\over{96}} + O(s(0)-1),\quad
s(0)-1\ll 1,$$
$$xg(x) = {\pi^6\over{960s(0)}} + {17\pi^8\over
{2^6 15 s(0)^2}} + O(s(0)^{-3}),\quad
s(0)>>1.\eqn(B.4)$$

\newpage

{\bf References }

\item{1.} S. E. Esipov, T. J. Newman, Phys.\ Rev.\ E, {\bf 48}, 1046, (1993).
\item{2.} J.M. Burgers, {\em The Non-linear Diffusion Equation }, Reidel,
Dordrecht (1974).
\item{3.} T. Gotoh and R.H. Kraichnan, Phys. Fluids A {\bf 5}(2), 445, (1993).
\item{4.} D. Forster, D.R. Nelson, and M.J. Stephen, Phys.\ Rev.\ A, {\bf 16},
732, (1977).
\item{5.} M. Kardar, G. Parisi and Y.-C. Zhang, Phys.\ Rev.\ Lett.,
{\bf 56}, 889 (1986).
\item{6.} S. N. Gurbatov, A. I. Saichev and I. G. Yakushkin, Sov.\ Phys.\ Usp.,
{\bf 26}(10), 857 (1983);
S. Gurbatov, A. Malakhov, and A. Saichev, {\it ``Nonlinear Random Waves and
and Turbulence in Nondispersive Media: Waves, Rays and Particles''},
Manchester,
Manchester University Press, 1991.
\item{7.} S.E. Esipov, Zh. Eksp. Teor. Fiz. {\bf 94}, 80, (1988), [Sov. Phys.
JETP, {\bf 67}, 257, (1988)].
\item{8.} E.D'Hoker and R.Jackiw, Phys.\ Rev.\ D, {\bf 26}, 3517, (1993).
\item{9.} T. Tatsumi and S. Kida,, J.\ Fluid.\ Mech., {\bf 55}, 659, (1972).
\item{10.} J.M. Kosterlitz and D.J. Thouless, J.\ Phys.\ C: Soild State Phys.,
{\bf 6}, 1181, (1973); J.M. Kosterlitz, J.\ Phys.\ C: Soild State Phys.,
{\bf 7}, 1046, (1974); J.V. Jose, L.P. Kadanoff, S. Kirkpatrick and
D.R. Nelson, Phys.\ Rev. B, {\bf 16}, 1217, (1977).
\item{11.} J. Zinn-Justin,
{\it Quantum Field Theory and Critical Phenomena}, Claredon Press, Oxford,
1990.
\item{12.} J. Krug and H. Spohn, ``{\it Kinetic roughening of growing
interfaces} in ``Solids far from Equilibrium'', ed. by C. Godreche, Cambridge
University Press, 1992.
\item{13.} A.S. Monin and A.M. Yaglom, {\it ``Statistical Fluid Mechanics''},
M.I.T., Cambridge, 1975.
\item{14.} R.P. Feynman and A.R. Hibbs,
{\it ``Quantum Mechanics and Path Integrals''}, McGraw-Hill, N.Y. 1965.
\item{15.} S.F. Edwards and D.R. Wilkinson, Proc. Roy. Soc. London A {\bf 381},
17, (1982)
\item{16.} This number is known with computer assistance.
Burgers obtained $\eta = 0.526$ (see formulas (44.7) and
(38.17a) in Ref.2), while we found by integrating Eq\main\ that
$\eta = 0.499$.
\item{17.} L.-H. Tang, T. Nattermann and B.M. Forrest, Phys.\ Rev.\ Lett.
{\bf 65}, 2422, (1990).

\chapter{Figure Captions}
\item{Fig.1} Log-log plot of $\langle{h}\rangle$ vs. time obtained by numerical
integration of Eq\main\ and using \heig, \psidef. The tilted line is
the Burgers asymptotic dependence
$\langle{h}\rangle = 3\eta D^{1/3} (\lambda{t})^{1/3}
$. The horizontal lines are the saturation heights, given by Eq\satur,
the corresponding heights used in simulation are $H = 3,\ 5,\ 10,\ 20$ from the
bottom to top.
Other parameters used: $D = \nu =
\lambda = 1$. Larger values of $H$ require too prolonged simulations.
\item{Fig.2} Log-log plot of kinetic energy $E$ vs. time obtained by
using $E = \lambda\partial_t\langle{h}\rangle$ and data of Fig.1.
The long tilted line is the Burgers asymptotic dependence, $E = \eta
 D^{1/3}
\lambda^{4/3} t^{-2/3}$. Other lines are the asymptotic dependences
given by Eq\fisum.
At these $H$ one can not use Eq\ajs\ yet, since the limit of stong turbulence
is not achieved. The transient $E\sim {t}^{-2}$ behavior just
starts to appear,
one may see only an
indication of this in the form of a kink-like bend on the curve with $H=20$.
\item{Fig.3} Log-log plot of $\langle{h}\rangle$ vs. time obtained by numerical
integration of Eq\maindiff\ and absorbing line \heighu (this is the
invisid limit). Solid lines
are the Burgers asymptotic dependence and saturation values $\langle{h}\rangle
= H$. The corresponding heights used in simulation are
$H = 3,\ 5,\ 10,\ 20$ from the bottom to top. Other parameters used: $D =
4, \lambda =1$.
\item{Fig.4} Log-log plot of kinetic energy $E$ vs. time obtained by
using $E = \lambda\partial_t\langle{h}\rangle$ and data of Fig.3.
The long tilted line is the Burgers asymptotic dependence,
other lines are the asymptotic dependences given by Eq\entrans.

\end